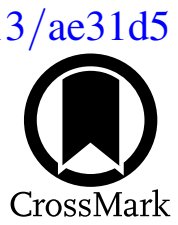

# Using Gravity as an Age Indicator for Young Stars

Michael Connelley[1,3], Christian Flores[2], and Bo Reipurth[1]
[1] University of Hawaii at Mānoa, 640 N. Aohoku Pl., Hilo, HI 96720, USA; mconnell@hawaii.edu
[2] Academia Sinica Institute of Astronomy and Astrophysics, 11F of Astronomy-Mathematics Building, AS/NTU, No. 1, Sec. 4, Roosevelt Rd., Taipei 10617, Taiwan

## Abstract

The age of a young star is difficult to measure and yet is a fundamental property that is essential for understanding stellar evolution. Accurate age indicators across the early phases of low-mass stars are scarce and often rely on properties of the circumstellar environment. We investigate stellar surface gravity as a practical age estimator from the class I to the class III phases of low-mass stars. Gravity increases monotonically as young stars descend their Hayashi tracks. Gravity is a physical property of the central star and is independent of the circumstellar environment or evolutionary models. Also, gravity can be measured for individual objects. We performed two tests for the reliability of gravity as an age indicator: (1) the relative ages of the components of young binary stars and (2) the relative ages of a young moving group. We show that young binary stars have consistent gravities, as expected, if they are coeval. Similarly, we find similar gravities among the TW Hya group and derive a median age of $12.8 \pm 1.2$ Myr. In both cases, differences in gravities may reflect real differences in the ages. We show that there is no correlation between veiling emission and gravity, showing that circumstellar disk emission does not affect our ability to infer ages from measurements of photospheric stellar gravity.

## 1. Introduction

### 1.1. Age Indicators for Young Stars

Because our goal is to study how stars form, it is important to put our observations into chronological order. Incorrect age estimates will drive us toward incorrect conclusions about the evolution of young stars, their disks, and their environment. As such, having accurate ages of young stars is very important

Several commonly used age indicators for young stars trace the infrared circumstellar environment and are unrelated to the central star itself. The spectral energy distribution slope is the basis for the class I, II, and III sequence (C. J. Lada 1987). This classification is purely empirical, and extinction from the envelope or an edge-on-disk can affect the spectral index. Infrared excess, accretion, and extinction are also used, but each is known to be variable (M. S. Connelley & T. P. Greene 2014). H$\alpha$ equivalent width is used to classify an object as a classical or weak-line T Tauri star. However, the H$\alpha$ equivalent width is variable, and objects near the threshold frequently cross back and forth between classical and weak-line T Tauri stars (C. Bertout & J. Bouvier 1989).

Ages of stars have also been estimated by placing them on a color–magnitude diagram with a comparison to theoretical evolutionary tracks. However, different theoretical tracks disagree on the physical properties of a star (i.e., age and mass) for a given set of observations. Comparison with evolutionary tracks typically requires determining the $T_{\rm eff}$ and the luminosity of the star. Colors can be used to estimate $T_{\rm eff}$, but both colors and luminosity are affected by extinction, which is often difficult to accurately measure. Furthermore, accretion adds to the luminosity of the star and can also affect the color. These effects are stronger for higher-accreting and more deeply embedded young stellar objects (YSOs), where the uncertainties involved in estimating the extinction and accretion effects can be substantial.

### 1.2. Gravity as an Age Indicator

The surface gravity of an isolated star depends only on its mass and radius via the relation $g = MR^{-2}$ (in solar units). C. Hayashi (1961) showed that fully convective stars evolve in the H-R diagram at a nearly constant temperature. Theoretical models (e.g., I. Baraffe et al. 2015; G. A. Feiden 2016) show that young low-mass ($M < 2\,M_\odot$) stars evolve down the Hayashi tracks with the surface gravity continuously increasing with age (above and to the right of the red dashed line in Figure 1, which shows where 60% of the radius of the star is radiative). Since the isochrones are roughly parallel with lines of constant gravity while stars are descending the Hayashi track, gravity is a nonlinear indicator of the relative stellar age. Isochrones become largely parallel to the mass tracks as stars evolve onto Henyey tracks, and gravity then loses most of its dependence on age. Gravity is a useful age indicator until stars evolve off of their Hayashi tracks.

While different models disagree on the conversion from gravity to age, and unknown factors in the initial conditions may add to the uncertainty, they do agree that gravity monotonically increases with time (Figure 2). Thus, while using gravity to establish absolute age is model dependent, gravity by itself is a good indicator of *relative* age. Gravity as an age indicator has the advantages that it is a physical property of the central star itself and is independent of circumstellar environment, distance, ability to measure luminosity or temperature, line-of-sight extinction, and modest accretion.

[3] Staff Astronomer at the Infrared Telescope Facility, which is operated by the University of Hawaii under contract 80HQTR24DA010 with the National Aeronautics and Space Administration.

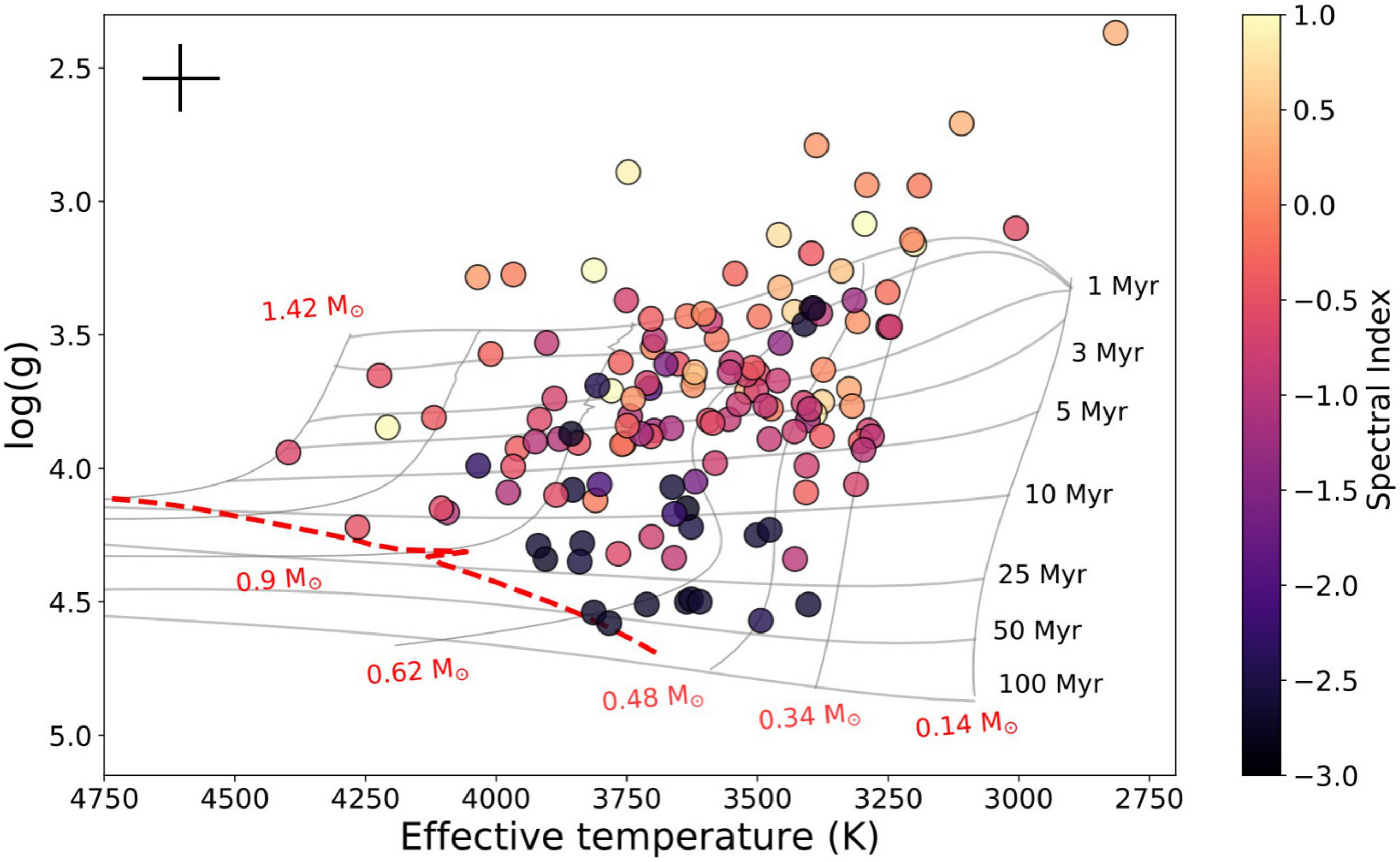

**Figure 1.** The G. A. Feiden (2016) models show that stars on the Hayashi tracks (above the red dashed line) evolve vertically downward on this plot because log($g$) primarily traces age while temperature primarily traces mass. Log($g$) is an effective age tracer while stars are on the Hayashi tracks where the isochrones and mass tracks are relatively orthogonal. Our median uncertainties (±75 K and ±0.13 dex log($g$)) are shown in the top left corner. The median age uncertainty for individual stars is ±45% of the age estimate (i.e., the median uncertainty for a 10 Myr age would be ±4.5 Myr). Our sample stars have log($g$) values from 2.5 to 4.5; hence, their inferred ages are distributed from <1 Myr to ∼100 Myr.

R. J. White et al. (2007) suggested using surface gravity to infer the ages of young stars, including embedded class I YSOs. M. Cottaar et al. (2014) used high-resolution $H$-band spectra to measure gravities of young stars and to estimate the ages of young star clusters. N. Da Rio et al. (2016) used similar methods to estimate the ages of young stars in the Orion star-forming region.

## 2. Sample and Observations

This Letter selects wide binaries and members of the TW Hya group, the sample in M. S. Connelley et al. (2026, in preparation). Observations are described in C. Flores et al. (2019, 2020, 2022) and in M. S. Connelley et al. (2026, in preparation). The observations were done using iSHELL (J. Rayner et al. 2022) on the Infrared Telescope Facility (IRTF) in the middle $K$-band (K2) mode (2.09–2.38 $\mu$m). Our study includes 106 targets.

The physical properties of the young stars are from C. Flores et al. (2022, 2024). The process of determining stellar parameters and their uncertainties from high-resolution spectra is detailed in C. Flores et al. (2019). Very briefly, they use MOOGstokes (C. P. Deen 2013) to generate model spectra including magnetic field effects. They then use Markov Chain Monte Carlo to simultaneously fit temperature, rotation, gravity, microturbulence, veiling, and the magnetic field strength. The properties of our sample are listed in Table 1.

### 2.1. Biases in Sample Selection

Our sample is limited to bright ($K < 11$) targets so that we can get a high signal-to-noise ratio (∼100 on the continuum) spectrum at high enough spectral resolution to resolve the line profiles. Also, the veiling needs to be low enough ($r_K < 3$) that the photospheric lines can be measured. While veiling was rarely an issue for the class II or III targets, only about half of class I YSOs have veiling low enough to show photospheric lines in their spectrum (M. S. Connelley & T. P. Greene 2010). Low-veiling class I YSOs have lower spectral index overall than high-veiling class I YSOs, and thus, low-veiling class I YSOs may be more evolved on average.

Most of the class II YSOs were selected to have kinematic masses from Atacama Large Millimeter/submillimeter Array (ALMA) observations so that we could test the masses predicted by evolutionary models (C. Flores et al. 2022). Our class II sample is thus biased in favor of YSOs with relatively large and massive disks, which, in turn, may impart a bias with regards to the age or mass of the central YSO. We have very few class II stars with spectral indices in the range from $-2 < \alpha < -1$, likely due to the selection of class II YSOs with ALMA resolved disks. To help fill in this gap, we included results from R. López-Valdivia et al. (2021), who performed an analysis similar to C. Flores et al. (2022) on their high-resolution near-IR spectra.

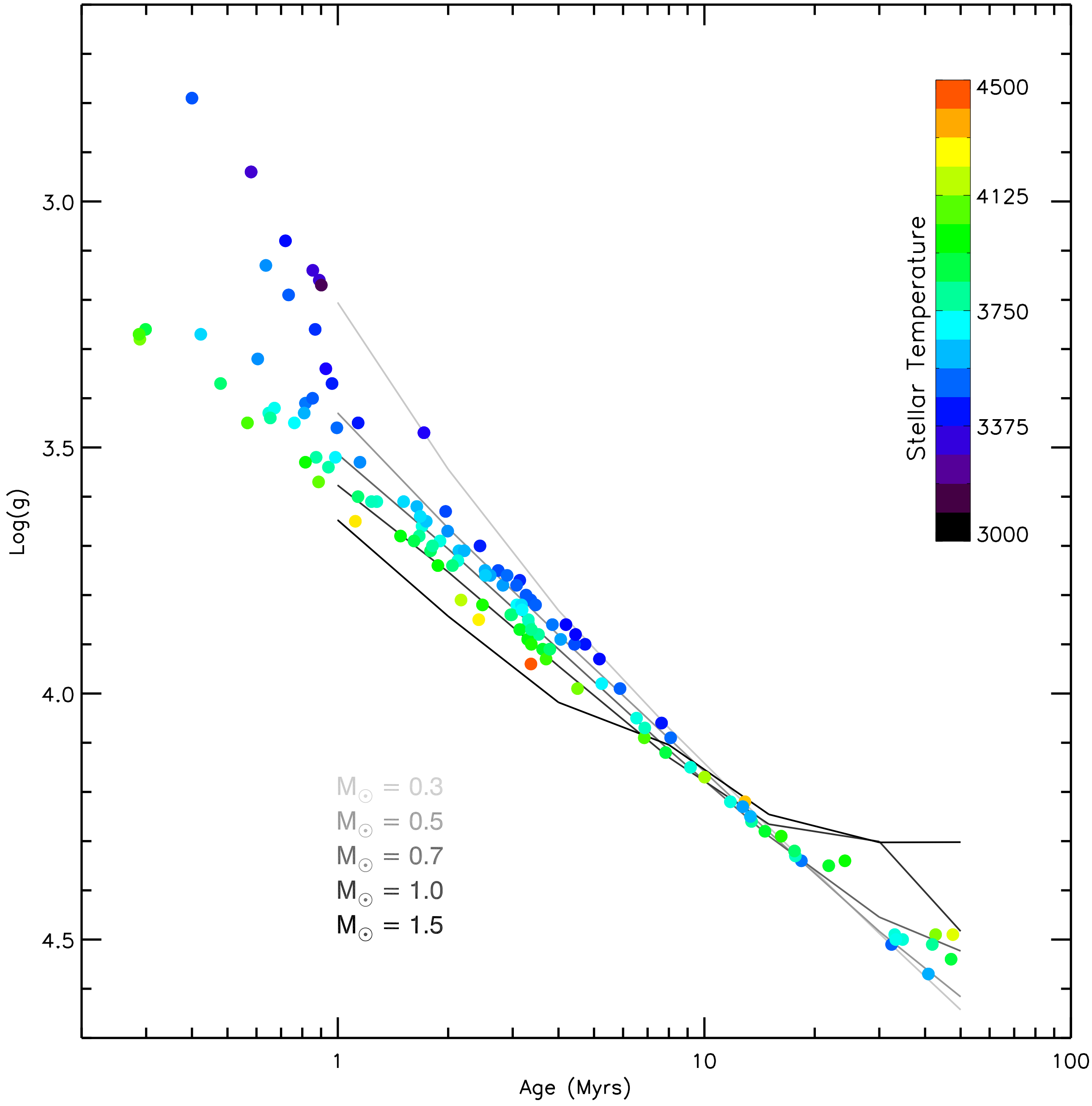

**Figure 2.** Overplotting our data on the age-versus-log($g$) relation shows how the cooler stars overlay the lower-mass tracks and hotter stars overlay the higher-mass tracks. There is some dispersion with temperature/mass for ages <5 Myr due to the tilt of the isochrones in Figure 1. For ages between 5 and 50 Myr, the isochrones are level in Figure 1, being relatively independent of temperature/mass, and at these ages, the mass tracks form a tight relationship between log($g$) and age.

## 3. Testing Gravity as an Age Indicator

Figure 1 shows that isochrones on a log($g$)-versus-temperature diagram are almost straight horizontal lines while stars are descending the Hayashi track. Figure 2 shows that the tight correlation between log($g$) and age has a slight temperature dependence at very young ages, which we can account for, since the spectral modeling also produces a temperature estimate. Here, we test gravity as an age indicator in two ways, first by comparing the gravities of the components of young binaries and then by looking at the gravities of members of a young moving group.

### *3.1. Do Binary Components Have the Same Gravity?*

Binary stars offer a stringent test of gravity as an age indicator. The individual components may have different magnetic fields, accretion, extinction, or circumstellar environment, yet their gravities ought to be consistent as they should have the same age. Previous work by R. J. White & A. M. Ghez (2001) and A. L. Kraus & L. A. Hillenbrand (2009) both found that binaries in the Taurus/Auriga star-forming region are more coeval than random sampling from the region. P. Hartigan & S. J. Kenyon (2003) estimated the ages of close binaries based on their location on the H-R

**Table 1**
Temperatures, Gravities, Veiling, and Ages for Our Sample of Young Stars

| Object | Temp. (K) | log(*g*) (cm s$^{-2}$) | Veiling ($r_K$) | Age (Myr) |
|---|---|---|---|---|
| 2MASSJ16430128-1754274 | $3711^{+12}_{-12}$ | $4.51^{+0.01}_{-0.01}$ | $0.05^{+0.01}_{-0.01}$ | $41.9^{+2.8}_{-2.7}$ |
| 2MASSJ19102820-2319486 | $3402^{+7}_{-6}$ | $4.51^{+0.01}_{-0.01}$ | $0.00^{+0.01}_{-0.00}$ | $32.4^{+1.1}_{-1.1}$ |
| 2MASSJ20013718-3313139 | $3635^{+10}_{-12}$ | $4.50^{+0.01}_{-0.00}$ | $0.04^{+0.03}_{-0.01}$ | $34.7^{+1.8}_{-0.0}$ |
| 2MASSJ20055640-3216591 | $3626^{+22}_{-35}$ | $4.49^{+0.01}_{-0.01}$ | $0.12^{+0.06}_{-0.04}$ | $33.1^{+1.7}_{-1.6}$ |
| 2MASSJ21100535-1919573 | $3610^{+7}_{-8}$ | $4.50^{+0.01}_{-0.00}$ | $0.06^{+0.00}_{-0.00}$ | $33.3^{+1.5}_{-0.0}$ |
| AA Tau | $3593^{+91}_{-91}$ | $3.82^{+0.14}_{-0.14}$ | $0.99^{+0.02}_{-0.01}$ | $3.1^{+0.9}_{-0.9}$ |
| BP Tau | $3702^{+91}_{-90}$ | $4.26^{+0.14}_{-0.14}$ | $1.36^{+0.02}_{-0.02}$ | $13.4^{+3.9}_{-3.0}$ |
| CI Tau | $3888^{+92}_{-92}$ | $3.74^{+0.14}_{-0.15}$ | $1.86^{+0.03}_{-0.02}$ | $1.9^{+0.9}_{-0.8}$ |
| CX Tau | $3495^{+90}_{-90}$ | $3.75^{+0.14}_{-0.14}$ | $0.26^{+0.00}_{-0.01}$ | $2.5^{+0.4}_{-0.6}$ |
| CY Tau | $3404^{+90}_{-90}$ | $3.82^{+0.14}_{-0.14}$ | $0.58^{+0.01}_{-0.01}$ | $3.5^{+0.4}_{-0.4}$ |
| DE Tau | $3460^{+91}_{-91}$ | $3.67^{+0.14}_{-0.14}$ | $1.15^{+0.02}_{-0.01}$ | $2.0^{+0.5}_{-0.4}$ |
| DG Tau | $4035^{+58}_{-70}$ | $3.28^{+0.13}_{-0.20}$ | $3.6^{+0.3}_{-0.38}$ | $0.29^{+0.19}_{-0.16}$ |
| DK Tau A | $3501^{+90}_{-90}$ | $3.71^{+0.14}_{-0.14}$ | $1.90^{+0.02}_{-0.03}$ | $2.2^{+0.6}_{-0.6}$ |
| DK Tau B | $3411^{+92}_{-90}$ | $3.76^{+0.14}_{-0.14}$ | $1.10^{+0.02}_{-0.02}$ | $2.9^{+0.6}_{-0.5}$ |
| DL Tau | $3917^{+91}_{-91}$ | $3.82^{+0.14}_{-0.14}$ | $2.97^{+0.02}_{-0.04}$ | $2.5^{+0.7}_{-0.6}$ |
| DM Tau | $3302^{+90}_{-90}$ | $3.90^{+0.14}_{-0.14}$ | $0.03^{+0.01}_{-0.00}$ | $4.7^{+0.3}_{-0.4}$ |
| DN Tau | $3521^{+91}_{-90}$ | $3.65^{+0.14}_{-0.14}$ | $0.62^{+0.01}_{-0.01}$ | $1.7^{+0.5}_{-0.3}$ |
| DoAr 25 | $3697^{+60}_{-60}$ | $3.52^{+0.14}_{-0.09}$ | $0.50^{+0.03}_{-0.03}$ | $0.87^{+0.41}_{-0.28}$ |
| DoAr 33 | $3697^{+22}_{-35}$ | $3.86^{+0.04}_{-0.05}$ | $0.69^{+0.02}_{-0.01}$ | $3.3^{+1.4}_{-1.0}$ |
| DoAr 43 | $4223^{+63}_{-56}$ | $3.65^{+0.13}_{-0.20}$ | $1.95^{+0.15}_{-0.15}$ | $1.1^{+0.5}_{-0.4}$ |
| DS Tau | $3629^{+69}_{-48}$ | $3.73^{+0.11}_{-0.10}$ | $1.66^{+0.01}_{-0.02}$ | $2.1^{+1.0}_{-0.6}$ |
| EC 92 | $3747^{+72}_{-78}$ | $2.89^{+0.12}_{-0.11}$ | $0.91^{+0.1}_{-0.1}$ | $0.09^{+0.05}_{-0.03}$ |
| EC 95 | $4207^{+116}_{-72}$ | $3.85^{+0.11}_{-0.19}$ | $0.95^{+0.11}_{-0.37}$ | $2.4^{+1.3}_{-1.2}$ |
| Elias 2–32 | $3387^{+52}_{-59}$ | $2.79^{+0.1}_{-0.12}$ | $0.48^{+0.06}_{-0.05}$ | $0.4^{+0.1}_{-0.1}$ |
| Elias 2–33 | $3290^{+88}_{-87}$ | $2.94^{+0.21}_{-0.11}$ | $1.50^{+0.13}_{-0.12}$ | $0.6^{+0.3}_{-0.1}$ |
| EPIC 211068400 | $4984^{+6}_{-104}$ | $4.41^{+0.03}_{-0.02}$ | $0.12^{+0.01}_{-0.03}$ | $37.7^{+17.9}_{-8.6}$ |
| FM Tau | $3381^{+32}_{-68}$ | $3.90^{+0.05}_{-0.05}$ | $1.67^{+0.03}_{-0.03}$ | $4.4^{+0.8}_{-0.6}$ |
| FP Tau | $3396^{+84}_{-63}$ | $3.40^{+0.20}_{-0.10}$ | $0.26^{+0.01}_{-0.01}$ | $0.9^{+0.7}_{-0.2}$ |
| FX Tau A | $3555^{+48}_{-56}$ | $3.82^{+0.07}_{-0.08}$ | $0.69^{+0.04}_{-0.03}$ | $3.2^{+0.8}_{-0.7}$ |
| GK Tau A | $3652^{+62}_{-56}$ | $3.61^{+0.12}_{-0.11}$ | $0.91^{+0.40}_{-0.39}$ | $1.3^{+0.7}_{-0.4}$ |
| GM Aur | $3842^{+41}_{-60}$ | $3.91^{+0.06}_{-0.09}$ | $0.59^{+0.02}_{-0.03}$ | $3.6^{+0.9}_{-1.0}$ |
| GO Tau | $3429^{+44}_{-33}$ | $3.86^{+0.05}_{-0.04}$ | $0.28^{+0.03}_{-0.03}$ | $3.8^{+0.6}_{-0.4}$ |
| GSS 26 | $3496^{+91}_{-81}$ | $3.43^{+0.23}_{-0.10}$ | $0.53^{+0.02}_{-0.04}$ | $0.8^{+0.3}_{-0.2}$ |
| GSS 37 W | $3590^{+32}_{-46}$ | $3.45^{+0.13}_{-0.06}$ | $0.48^{+0.02}_{-0.03}$ | $0.8^{+0.3}_{-0.2}$ |
| GSS 39 | $3397^{+60}_{-77}$ | $3.19^{+0.10}_{-0.14}$ | $1.82^{+0.07}_{-0.07}$ | $0.7^{+0.2}_{-0.2}$ |
| GV Tau S | $3812^{+66}_{-74}$ | $3.26^{+0.12}_{-0.17}$ | $2.24^{+0.15}_{-0.18}$ | $0.30^{+0.18}_{-0.15}$ |
| [*GY*92] 33 | $3755^{+55}_{-63}$ | $3.91^{+0.07}_{-0.1}$ | $0.29^{+0.02}_{-0.02}$ | $3.8^{+1.1}_{-1.1}$ |
| [*GY*92] 235 | $3374^{+88}_{-162}$ | $3.63^{+0.17}_{-0.34}$ | $2.21^{+0.18}_{-0.17}$ | $2.0^{+1.2}_{-1.2}$ |
| [*GY*92] 284 | $3390^{+64}_{-65}$ | $3.81^{+0.09}_{-0.11}$ | $0.46^{+0.06}_{-0.04}$ | $3.3^{+1.3}_{-0.9}$ |
| Haro 1–16 | $3959^{+113}_{-126}$ | $3.93^{+0.09}_{-0.13}$ | $2.32^{+0.16}_{-0.16}$ | $3.7^{+1.6}_{-1.1}$ |
| Haro 6–13 | $3700^{+96}_{-109}$ | $3.54^{+0.21}_{-0.25}$ | $1.55^{+0.12}_{-0.13}$ | $0.94^{+1.2}_{-0.58}$ |
| Haro 6–33 | $3623^{+61}_{-53}$ | $3.69^{+0.12}_{-0.14}$ | $0.55^{+0.03}_{-0.03}$ | $1.9^{+1.1}_{-0.8}$ |
| Haro 6–28 | $3287^{+31}_{-39}$ | $3.86^{+0.05}_{-0.06}$ | $0.30^{+0.03}_{-0.03}$ | $4.2^{+0.68}_{-0.69}$ |
| HIP110526E | $3677^{+15}_{-19}$ | $5.00^{+0.03}_{-0.03}$ | $0.11^{+0.01}_{-0.02}$ | Red Line |
| HIP110526W | $3751^{+17}_{-17}$ | $5.16^{+0.03}_{-0.03}$ | $0.00^{+0.01}_{-0.00}$ | Red Line |
| HIP113579 | $6597^{+153}_{-88}$ | $3.85^{+0.15}_{-0.18}$ | $0.01^{+0.06}_{-0.03}$ | Red Line |
| HIP11437NW | $3786^{+12}_{-12}$ | $4.99^{+0.01}_{-0.02}$ | $0.08^{+0.01}_{-0.01}$ | Red Line |
| HIP11437SE | $4048^{+39}_{-96}$ | $4.49^{+0.01}_{-0.06}$ | $-0.03^{+0.08}_{-0.02}$ | $42.7^{+3.8}_{-17.2}$ |
| HIP12545 | $4157^{+118}_{-47}$ | $4.49^{+0.04}_{-0.02}$ | $-0.11^{+0.02}_{-0.01}$ | $47.6^{+22.9}_{-4.5}$ |
| HIP12635 | $4351^{+491}_{-239}$ | $4.49^{+0.31}_{-0.04}$ | $-0.04^{+0.06}_{-0.05}$ | Red Line |
| HIP16563NW | $5165^{+64}_{-70}$ | $4.75^{+0.03}_{-0.06}$ | $-0.04^{+0.07}_{-0.01}$ | Red Line |
| HIP16563SE | $3813^{+11}_{-11}$ | $4.54^{+0.02}_{-0.03}$ | $-0.01^{+0.00}_{-0.01}$ | $47.1^{+5.9}_{-7.6}$ |
| HIP17695 | $3673^{+14}_{-13}$ | $5.02^{+0.03}_{-0.02}$ | $0.01^{+0.00}_{-0.01}$ | Red Line |

**Table 1**
(Continued)

| Object | Temp. (K) | log(*g*) (cm s$^{-2}$) | Veiling ($r_K$) | Age (Myr) |
|---|---|---|---|---|
| HIP18859 | $6731^{+111}_{-101}$ | $4.67^{+0.12}_{-0.12}$ | $0.37^{+0.05}_{-0.05}$ | Red Line |
| HIP81084 | $3784^{+11}_{-10}$ | $4.58^{+0.02}_{-0.02}$ | $0.00^{+0.01}_{-0.00}$ | Red Line |
| HIP82688 | $5855^{+68}_{-62}$ | $4.23^{+0.06}_{-0.08}$ | $0.07^{+0.02}_{-0.03}$ | Red Line |
| HK Tau A | $3509^{+49}_{-45}$ | $3.62^{+0.09}_{-0.10}$ | $0.74^{+0.03}_{-0.03}$ | $1.6^{+0.6}_{-0.5}$ |
| HO Tau | $3428^{+48}_{-46}$ | $4.34^{+0.07}_{-0.09}$ | $1.08^{+0.06}_{-0.06}$ | $18.4^{+4.8}_{-4.8}$ |
| HP Tau | $4118^{+40}_{-40}$ | $3.81^{+0.09}_{-0.10}$ | $1.80^{+0.08}_{-0.08}$ | $2.2^{+0.9}_{-0.7}$ |
| IP Tau | $3659^{+36}_{-41}$ | $4.33^{+0.05}_{-0.04}$ | $1.24^{+0.05}_{-0.04}$ | $17.7^{+4.0}_{-2.7}$ |
| IQ Tau | $3549^{+71}_{-51}$ | $3.61^{+0.11}_{-0.12}$ | $0.91^{+0.06}_{-0.04}$ | $1.5^{+0.7}_{-0.5}$ |
| IRAS 03260+3111 B | $3200^{+90}_{-88}$ | $3.16^{+0.16}_{-0.16}$ | $0.42^{+0.05}_{-0.06}$ | $0.89^{+0.34}_{-0.25}$ |
| IRAS 03301+3111 | $3457^{+73}_{-85}$ | $3.32^{+0.13}_{-0.14}$ | $1.27^{+0.08}_{-0.06}$ | $0.60^{+0.33}_{-0.23}$ |
| IRAS 04108+2803 E | $3779^{+65}_{-84}$ | $3.71^{+0.13}_{-0.16}$ | $0.89^{+0.05}_{-0.06}$ | $1.8^{+1.2}_{-0.8}$ |
| IRAS 04113+2758 S | $3204^{+41}_{-16}$ | $3.14^{+0.06}_{-0.05}$ | $1.02^{+0.02}_{-0.03}$ | $0.85^{+0.11}_{-0.08}$ |
| IRAS 04181+2665 M | $3520^{+68}_{-45}$ | $3.71^{+0.11}_{-0.12}$ | $0.21^{+0.03}_{-0.03}$ | $2.1^{+0.95}_{-0.71}$ |
| IRAS 04181+2665 S | $3376^{+84}_{-128}$ | $3.75^{+0.13}_{-0.36}$ | $1.98^{+0.18}_{-0.19}$ | $2.7^{+1.4}_{-1.8}$ |
| IRAS 04295+2251 | $3429^{+136}_{-134}$ | $3.41^{+0.33}_{-0.22}$ | $1.98^{+0.17}_{-0.18}$ | $0.82^{+1.6}_{-0.42}$ |
| IRAS 04489+3042 | $3324^{+73}_{-210}$ | $3.70^{+0.12}_{-0.47}$ | $1.58^{+0.13}_{-0.10}$ | $2.4^{+1.0}_{-1.8}$ |
| IRAS 04591−0856 | $3339^{+129}_{-122}$ | $3.26^{+0.18}_{-0.22}$ | $0.71^{+0.1}_{-0.09}$ | $0.87^{+0.46}_{-0.35}$ |
| IRAS 05379−0758(2) | $3459^{+97}_{-94}$ | $3.13^{+0.17}_{-0.12}$ | $0.58^{+0.06}_{-0.06}$ | $0.64^{+0.32}_{-0.16}$ |
| IRAS 05555−1405(4) | $3966^{+155}_{-164}$ | $3.27^{+0.53}_{-0.27}$ | $0.87^{+0.43}_{-0.27}$ | $0.29^{+2.0}_{-0.19}$ |
| IRAS 16285−2358 | $3328^{+80}_{-79}$ | $3.77^{+0.16}_{-0.15}$ | $1.69^{+0.13}_{-0.11}$ | $3.1^{+1.8}_{-1.1}$ |
| IRAS 16288−2450 W2 | $3473^{+41}_{-70}$ | $3.78^{+0.07}_{-0.10}$ | $0.41^{+0.03}_{-0.04}$ | $2.8^{+0.71}_{-0.78}$ |
| IRAS 19247+2238(1) | $3634^{+79}_{-65}$ | $3.43^{+0.17}_{-0.11}$ | $2.00^{+0.08}_{-0.08}$ | $0.65^{+0.59}_{-0.22}$ |
| IRAS 19247+2238(2) | $3577^{+99}_{-87}$ | $3.52^{+0.17}_{-0.11}$ | $1.41^{+0.53}_{-0.15}$ | $0.99^{+0.86}_{-0.33}$ |
| LkCa 15 | $4093^{+41}_{-42}$ | $4.17^{+0.08}_{-0.17}$ | $1.07^{+0.05}_{-0.06}$ | $10.0^{+6.2}_{-6.4}$ |
| [*TS*84] IRS5 N | $3295^{+89}_{-81}$ | $3.08^{+0.16}_{-0.08}$ | $0.38^{+0.05}_{-0.04}$ | $0.72^{+0.28}_{-0.11}$ |
| ROX 25 | $4397^{+52}_{-50}$ | $3.94^{+0.20}_{-0.13}$ | $1.04^{+0.18}_{-0.14}$ | $3.4^{+3.9}_{-1.3}$ |
| ROX 27 | $3761^{+45}_{-55}$ | $3.60^{+0.10}_{-0.14}$ | $1.28^{+0.03}_{-0.05}$ | $1.1^{+0.5}_{-0.4}$ |
| SR 24 S | $4010^{+122}_{-141}$ | $3.57^{+0.32}_{-0.46}$ | $3.03^{+0.53}_{-0.65}$ | $0.89^{+2.2}_{-0.73}$ |
| T Tau N | $3976^{+90}_{-90}$ | $3.45^{+0.14}_{-0.14}$ | $3.00^{+0.04}_{-0.04}$ | $0.57^{+0.41}_{-0.24}$ |
| TWA 3B | $3312^{+32}_{-14}$ | $4.06^{+0.08}_{-0.04}$ | $-0.25^{+0.01}_{-0.02}$ | $7.6^{+2.1}_{-0.9}$ |
| TWA 4 | $3766^{+150}_{-11}$ | $4.32^{+0.14}_{-0.02}$ | $0.12^{+0.01}_{-0.11}$ | $17.6^{+18.8}_{-1.7}$ |
| TWA 6A | $3919^{+13}_{-114}$ | $4.29^{+0.03}_{-0.02}$ | $-0.17^{+0.06}_{-0.01}$ | $16.2^{+4.4}_{-2.4}$ |
| TWA 7 | $3494^{+20}_{-18}$ | $4.57^{+0.02}_{-0.01}$ | $-0.05^{+0.02}_{-0.02}$ | $40.8^{+3.1}_{-2.9}$ |
| TWA 8A | $3691^{+28}_{-32}$ | $4.94^{+0.04}_{-0.04}$ | $-0.08^{+0.02}_{-0.03}$ | Red Line |
| TWA 9A | $3835^{+35}_{-45}$ | $4.28^{+0.06}_{-0.04}$ | $-0.01^{+0.04}_{-0.04}$ | $14.6^{+3.8}_{-2.1}$ |
| TWA 9B | $3476^{+37}_{-26}$ | $4.23^{+0.06}_{-0.04}$ | $-0.05^{+0.11}_{-0.13}$ | $12.7^{+2.8}_{-1.6}$ |
| TWA 13 N | $3637^{+14}_{-20}$ | $4.22^{+0.05}_{-0.05}$ | $-0.13^{+0.11}_{-0.01}$ | $11.8^{+2.3}_{-1.9}$ |
| TWA 13 S | $3635^{+16}_{-13}$ | $4.15^{+0.03}_{-0.03}$ | $-0.11^{+0.02}_{-0.01}$ | $9.2^{+1.0}_{-0.9}$ |
| TWA 23 | $3501^{+15}_{-17}$ | $4.25^{+0.05}_{-0.07}$ | $-0.04^{+0.02}_{-0.05}$ | $13.4^{+2.4}_{-2.8}$ |
| TYC 915-1391-1 | $3410^{+25}_{-21}$ | $3.46^{+0.08}_{-0.05}$ | $-0.16^{+0.02}_{-0.01}$ | $0.99^{+0.30}_{-0.15}$ |
| TYC 6349-0200-1 NW | $3906^{+106}_{-34}$ | $4.34^{+0.12}_{-0.01}$ | $0.08^{+0.02}_{-0.06}$ | $24.2^{+39.0}_{-1.9}$ |
| TYC 6349-0200-1 SE | $3610^{+22}_{-14}$ | $4.88^{+0.03}_{-0.03}$ | $-0.22^{+0.02}_{-0.06}$ | Red Line |
| TYC 6878-0195-1 | $3840^{+101}_{-37}$ | $4.35^{+0.04}_{-0.06}$ | $0.07^{+0.02}_{-0.09}$ | $21.8^{+5.8}_{-6.5}$ |
| UY Aur NE | $3602^{+45}_{-58}$ | $3.42^{+0.08}_{-0.11}$ | $1.23^{+0.05}_{-0.04}$ | $0.7^{+0.2}_{-0.2}$ |
| V347 Aur | $3190^{+105}_{-93}$ | $2.94^{+0.09}_{-0.08}$ | $1.17^{+0.09}_{-0.08}$ | $0.58^{+0.39}_{-0.15}$ |
| V710 Tau N | $3484^{+20}_{-34}$ | $3.76^{+0.04}_{-0.05}$ | $1.99^{+0.46}_{-0.41}$ | $2.6^{+0.4}_{-0.4}$ |
| V710 Tau S | $3400^{+35}_{-30}$ | $3.78^{+0.03}_{-0.03}$ | $0.22^{+0.01}_{-0.02}$ | $3.1^{+0.3}_{-0.3}$ |
| WSB 82 | $3879^{+135}_{-165}$ | $3.68^{+0.22}_{-0.30}$ | $3.49^{+0.46}_{-0.52}$ | $1.5^{+0.7}_{-0.5}$ |
| WL 20 E | $3621^{+80}_{-49}$ | $3.66^{+0.12}_{-0.12}$ | $0.32^{+0.03}_{-0.02}$ | $1.7^{+1.0}_{-0.62}$ |
| WL 20 W | $3390^{+53}_{-52}$ | $3.80^{+0.10}_{-0.11}$ | $0.14^{+0.03}_{-0.02}$ | $3.3^{+1.1}_{-0.91}$ |
| WLY2-42 | $3308^{+67}_{-46}$ | $3.45^{+0.17}_{-0.07}$ | $0.10^{+0.02}_{-0.03}$ | $1.1^{+0.72}_{-0.21}$ |
| YLW 58 | $3022^{+115}_{-22}$ | $3.17^{+0.31}_{-0.17}$ | $1.69^{+0.13}_{-0.15}$ | $0.9^{+1.2}_{-0.3}$ |

**Note.** Ages are calculated using the G. A. Feiden (2016) models. "Red line" designates stars where log(*g*) or temperature is too high for an age to be reliably estimated and are below the red line in Figure 1.

(This table is available in machine-readable form in the online article.)

**Table 2**
Binaries: Parameters from Table 1 Are Collected Here When We Have Resolved Spectra of Both Binary Components

| Object | Temp. (K) | log(g) (cm s$^{-1}$) | Age (Myr) |
|---|---|---|---|
| DK Tau A | $3501^{+46}_{-27}$ | $3.71^{+0.07}_{-0.09}$ | $2.2^{+0.6}_{-0.5}$ |
| DK Tau B | $3411^{+64}_{-17}$ | $3.76^{+0.06}_{-0.06}$ | $2.9^{+0.6}_{-0.5}$ |
| EC 92 | $3747^{+72}_{-78}$ | $2.89^{+0.12}_{-0.11}$ | <0.5 |
| EC 95 | $4207^{+116}_{-72}$ | $3.85^{+0.11}_{-0.19}$ | $2.4^{+1.3}_{-1.2}$ |
| Elias 2–32 | $3388^{+47}_{-62}$ | $2.79^{+0.10}_{-0.12}$ | <0.5 |
| Elias 2–33 | $3294^{+50}_{-74}$ | $2.94^{+0.21}_{-0.11}$ | <0.5 |
| IRAS 04181+2665 M | $3520^{+68}_{-45}$ | $3.71^{+0.11}_{-0.12}$ | $2.1^{+1.0}_{-0.7}$ |
| IRAS 04181+2665 S | $3376^{+84}_{-128}$ | $3.75^{+0.13}_{-0.36}$ | $2.7^{+1.4}_{-1.8}$ |
| IRAS 19247+2238(1) | $3634^{+79}_{-65}$ | $3.43^{+0.17}_{-0.11}$ | $0.6^{+0.6}_{-0.2}$ |
| IRAS 19247+2238(2) | $3577^{+99}_{-87}$ | $3.52^{+0.17}_{-0.11}$ | $1.0^{+0.9}_{-0.3}$ |
| TWA 9 A | $3835^{+35}_{-45}$ | $4.28^{+0.06}_{-0.04}$ | $14.6^{+3.8}_{-2.1}$ |
| TWA 9 B | $3476^{+37}_{-26}$ | $4.23^{+0.06}_{-0.04}$ | $12.7^{+2.2}_{-1.9}$ |
| TWA 13 N | $3637^{+14}_{-20}$ | $4.22^{+0.05}_{-0.05}$ | $11.8^{+2.2}_{-1.9}$ |
| TWA 13 S | $3635^{+16}_{-13}$ | $4.15^{+0.03}_{-0.03}$ | $9.2^{+1.0}_{-0.9}$ |
| V710 Tau N | $3484^{+20}_{-34}$ | $3.76^{+0.04}_{-0.05}$ | $2.6^{+0.4}_{-0.4}$ |
| V710 Tau S | $3400^{+35}_{-30}$ | $3.78^{+0.03}_{-0.03}$ | $3.1^{+0.3}_{-0.3}$ |
| WL 20 E | $3621^{+80}_{-49}$ | $3.66^{+0.12}_{-0.12}$ | $1.7^{+1.0}_{-0.6}$ |
| WL 20 W | $3390^{+53}_{-52}$ | $3.80^{+0.10}_{-0.11}$ | $3.3^{+1.1}_{-0.9}$ |

diagram and found that the ages of primary and secondary stars correlate better than random pairing. S. Correia et al. (2013) similarly found that visual binaries in the Orion Nebula Cluster were also more coeval than random sampling from the region. In a study of seven pre-main-sequence eclipsing binaries, M. Simon & J. Toraskar (2017) found that five out of seven were formed within 0.3 Myr of each other.

If log($g$) corresponds to age, the components of a binary ought to have similar gravities even if other properties are different. Even if the model-based conversion from log($g$) to age may be uncertain, the gravities of the stars in a binary ought to be similar. Our sample has nine binary pairs where we were able to extract the physical parameters for both binary components, which are listed in Table 2. Eight of these binary pairs have gravities and ages that are consistent with each other within the errors. The median difference in the log($g$) values is 0.07 dex, slightly less than the median uncertainty of 0.11 dex, so most of the binary components have log($g$) values similar to their companions. The median difference in age is 14% when neither age is an upper limit. The EC 92/95 is the only pair that has inconsistent gravity among the binaries and may not have formed together. However, excluding this pair does not change the median difference in log($g$) values for these binaries. In most (eight out of nine) of the cases we examined, the error bars on log($g$) (and hence age) overlap, showing that gravity does work as an age indicator.

While the error bars on log($g$) mostly overlap, EC 92/95 is the exception where they do not. EC 92/95 are two members of a dense group, and while they are currently found close to each other on the sky, they may have formed independently as there are other young stars nearby. Given that these stars are extremely young (A. A. Kaas et al. 2004; K. E. J. Haisch et al. 2006), it is possible that we could be seeing real age differences between these two stars.

Since the log($g$) measurement is based on the line profile, we do not expect modest accretion or extinction to affect the stellar surface gravity measurements as they only affect the continuum flux. Veiling and magnetic fields have negligible covariance with log($g$) (C. Flores et al. 2024) and similarly are not expected to affect our gravity measurements. DK Tau A and B have different infrared veiling, as do the components of IRAS 04181+2665 M/S. The components of DK Tau A/B and IRAS 04181+2665 M/S have different extinctions (M. S. Connelley et al. 2026, in preparation), while DK Tau A/B and V710 Tau N/S have different magnetic field strengths (C. Flores et al. 2022). Nevertheless, the binary components have similar inferred ages despite these differences in magnetic field strength, extinction, accretion, and veiling.

### 3.2. Age of the TW Hya Group

Our sample includes 10 stars from the TW Hya group. As they are expected to have similar ages, they should have similar gravities. Figure 3 shows the clustering of our sample of TW Hydra around the same log($g$). The mean log($g$) for our stars in the TW Hya group is $4.33 \pm 0.08$. This uncertainty is larger than the mean of the individual log($g$) uncertainties, largely due to two outliers: TWA 7 (log($g$) = 4.57) and TWA 8A (log($g$) = 4.94). If these are excluded, then the mean log($g$) for our stars in the TW Hya group is $4.23 \pm 0.03$, and the uncertainty on the mean is less than the individual uncertainties. Most (8 out of 10) of our sample of TW Hya members have similar gravities.

Several previous studies have estimated the age of the TW Hya group, including C. P. M. Bell et al. (2015; $10 \pm 3$ Myr), E. E. Mamajek (2005; >10 Myr), and K. Ujjwal et al. (2020; 6.5 Myr). C. Flores et al. (2022) showed that the G. A. Feiden (2016) models yielded masses that best matched ALMA dynamical masses. Using the G. A. Feiden (2016) models, the mean age of our sample of TW Hya members is $16.0 \pm 3.3$ Myr. Excluding TWA 7 and 8A (estimated to have ages of $41 \pm 3$ and >100 Myr, respectively), the mean age is $12.8 \pm 1.2$ Myr. We derive a similar age ($9.5 \pm 0.7$ Myr) using the I. Baraffe et al. (2015) evolutionary tracks. Our gravity measurements and the G. A. Feiden (2016) models yield ages consistent with prior age estimates derived using different methods. Excluding TWA 7 and 8A, our inferred ages for TWA members range from 7.6 to 17.6 Myr. As the median uncertainty of the ages is 2.4 Myr, this spread in ages may reflect real differences in the ages of the TWA members.

TWA 7 and TWA 8A are outliers with higher log($g$) than the rest of the TWA targets. We do not have an explanation as to why TWA 7 and 8 seem to be much older than other stars in the group. Their inferred ages are 40.8 and >100 Myr, respectively, whereas the rest of the group has ages from 7.6 to 17.6 Myr. Their stellar parameters are similar to the rest of the group except for $v\sin(i)$. TWA 7 and 8A are on the near side of the group, but there are other members of the group at similar distances, so the spatial location of these two stars is not exceptional. The $v\sin(i)$ of TWA 7 and 8A is much slower (~2 km s$^{-1}$) than the 13.3 km s$^{-1}$ median $v\sin(i)$ of the rest of the group. It is expected for stellar rotation to slow as stars age.

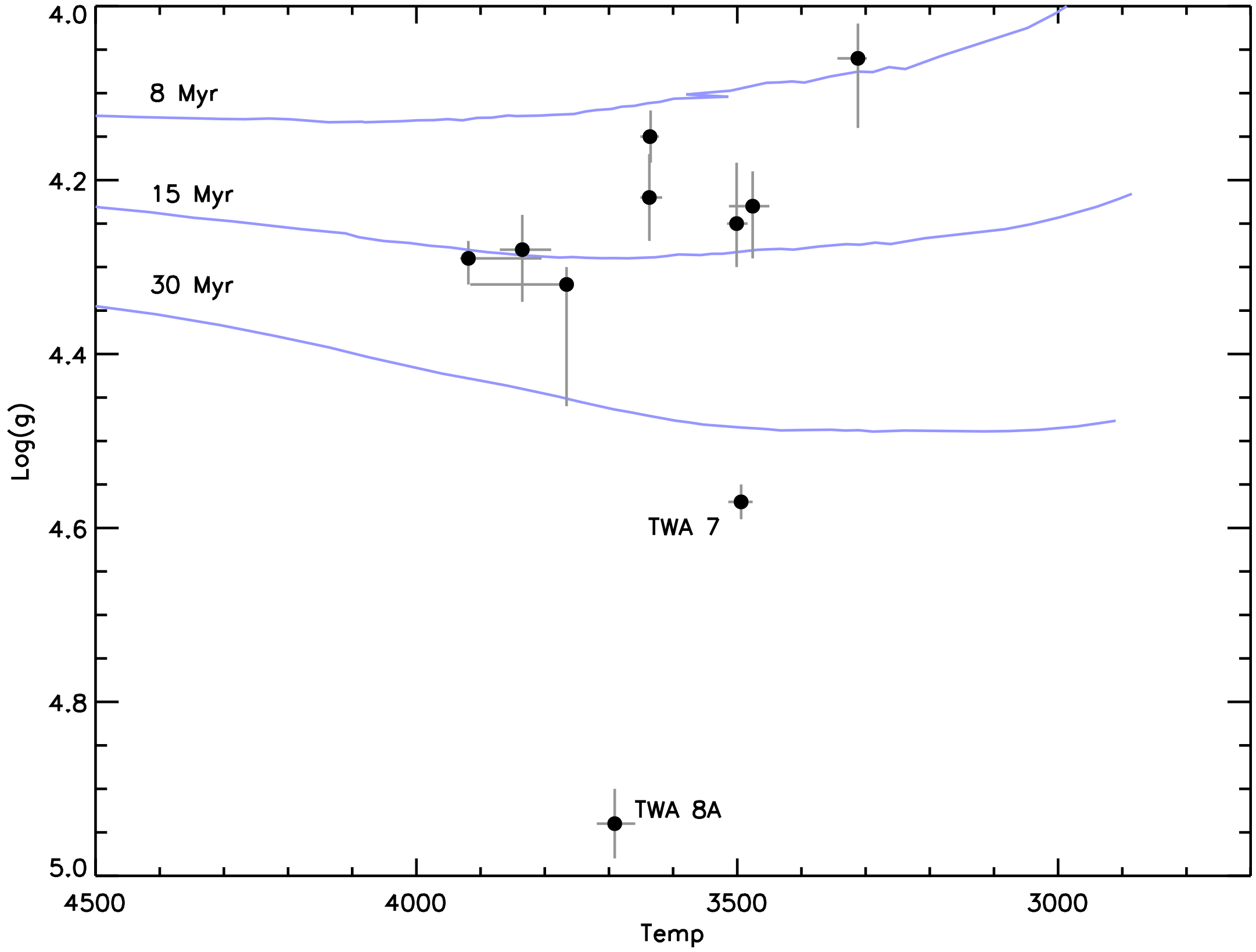

**Figure 3.** On the log($g$)-versus-temperature diagram, the members of the TW Hydra group are clustered around 13 Myr of age. G. A. Feiden (2016) isochrones are overlaid. The data suggest that the more massive (hotter) stars may have been born first.

Analysis of the ages of the TW Hydra members and young binaries supports our finding that log($g$) can accurately trace the relative ages of our sample of YSOs.

### *3.3. Can Veiling Affect Gravity Measurements?*

As the circumstellar flux contribution can be stronger than the light from the stellar photosphere at near-IR wavelengths, it is important to consider whether this contribution can affect our measurements of the stellar properties, and especially gravities, from the photospheric lines. Sources of this veiling that may contribute at 2 $\mu$m include the hot accretion footprint on the stellar photosphere, thermal dust emission from the inner edge of the disk, and gas inside of the disk inner edge.

We stress that our measurements of the stellar parameters rely on the photospheric line widths and relative strengths, whereas veiling uniformly affects the strengths of all lines within the limited bandwidth of our observations. We allow the code to adjust the veiling while optimizing the fit of the model spectrum to the data. The measurements of gravity come primarily from the line widths and line profiles, and so to affect our gravity measurements, the veiling would need to impact the apparent line widths. Sources of continuum emission, either from the accretion footprint on the star or thermal emission from dust, would not affect the line widths.

We estimate the veiling temperature while also varying the extinction and amount of veiling during our fitting of the 0.7–2.4 $\mu$m spectra from SpeX (J. T. Rayner et al. 2003), the low- to medium-resolution near-IR spectrograph on IRTF. Veiling is modeled as a single-temperature blackbody, where the temperature and the amount of veiling ($r_K$) are varied. The typical veiling temperature is in the 1000–1500 K range (M. S. Connelley & T. P. Greene 2010), which is consistent with the expected dust sublimation temperature at the disk inner edge.

We can also empirically investigate whether increased disk emission, traced by veiling, has any impact on our gravity measurements. Figure 4 shows that there is no correlation between $r_K$ and our gravity measurements; instead, we see a wide range of veiling among low-veiling class I and class II YSOs, whereas the class III YSOs have negligible veiling. If Figure 4 showed a correlation between veiling and gravity, we could not differentiate between a scenario where our gravity measurements is affected by veiling or genuine evolution of veiling with time. However, when considering high-veiling ($r_K > 1$) to low-veiling ($r_K < 1$) class I and II objects, the median log($g$) values are $3.71 \pm 0.06$ and $3.65 \pm 0.06$, respectively. The difference is much less than the 0.32 standard deviation in the log($g$) measurements. It is highly unlikely that changes in veiling versus time would exactly compensate for a possible effect of veiling on the line widths used to measure gravity. We conclude that disk emission has a negligible impact on our measurements of stellar physical parameters in general, and of gravity in particular, because (1) there is no empirical trend in the veiling with log($g$), (2) continuum flux sources cannot affect line widths, (3) we do not expect nor see absorption lines originating in the inner disk, and (4) we measure similar gravity for binary members with different amounts of veiling.

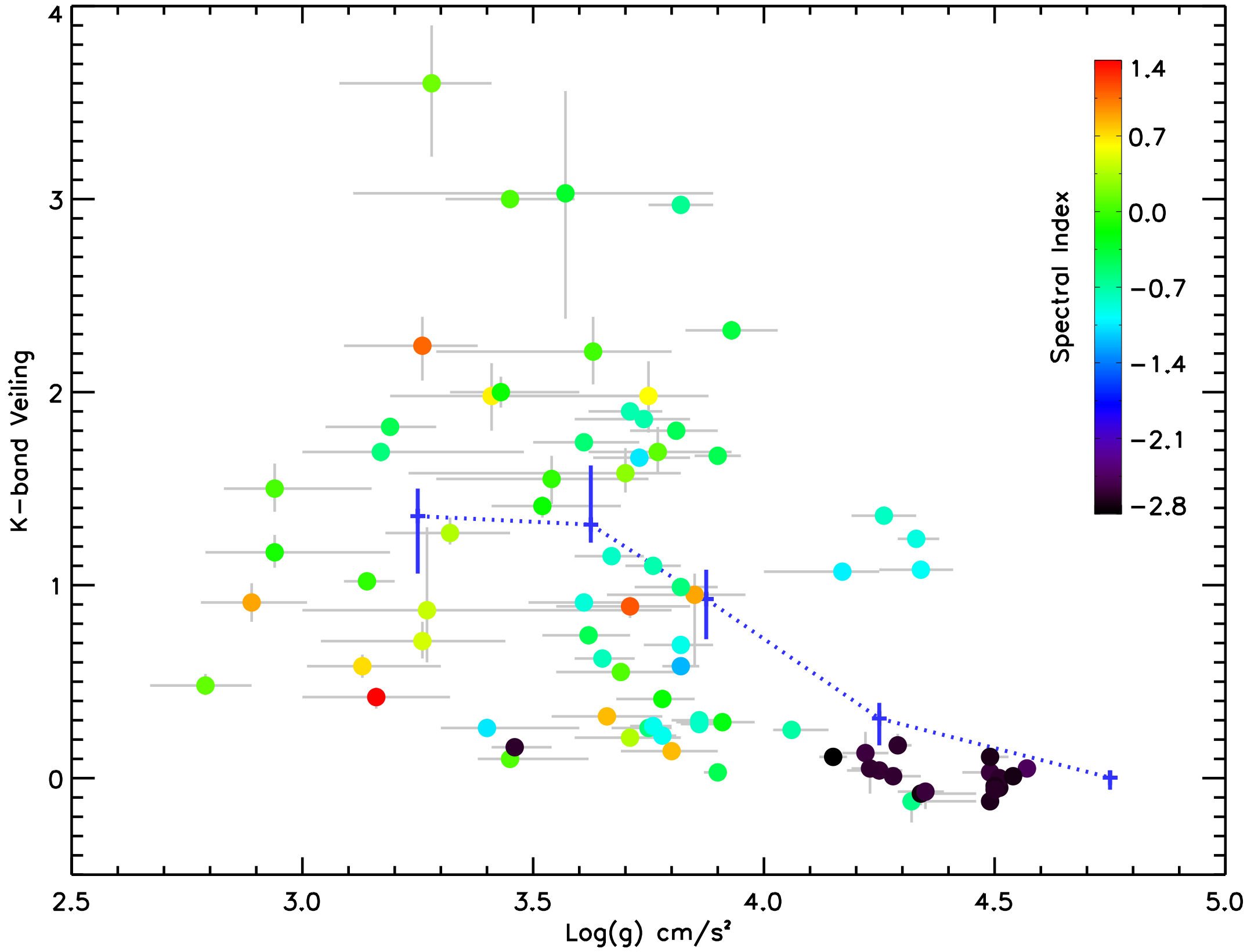

**Figure 4.** The plot of log($g$) versus veiling does not show a clear trend. For log($g$)$\lesssim$ 4 (age $\lesssim$ 10 Myr), stars have a wide range of veiling values with no clear trend. The class III stars have low veiling, as expected. For some stars, the veiling measurement is negative since higher metallicity increases line depths and thus acts like negative veiling. The blue line is the mean veiling for five bins of log($g$), with the uncertainties of the means shown as the error bars. The color bar reflects the spectral index of the stars.

## 4. Conclusions

We have explored using gravity as a tracer of age for young stars descending the Hayashi track. We primarily consider the relative ages of nine young binary pairs and eight members of the TW Hya group. Our key findings are the following:

(1) Binaries should have similar gravities if they are coeval, and indeed the gravities of the binary components are consistent with each other with a few exceptions. Our sample includes nine binary pairs, and their ages are consistent within the uncertainties for eight of them. The only outlier is EC 92/95, which is part of a young compact group and may have formed independently.

(2) We infer an age for the TW Hya group of $12.8 \pm 1.2$ Myr by using the gravities of eight members (excluding two outliers) and using the G. A. Feiden (2016) models. This age is broadly consistent with previous measurements. We derive a similar age using the I. Baraffe et al. (2015) models.

(3) Veiling does not affect gravity measurements and the ages that we infer. We see no trend in the log($g$) or inferred age with $K$-band veiling.

The next steps can include expanding to higher- and lower-mass stars to determine if the gravity–age relation holds. It would be valuable to be able to measure the ages of high-mass stars such as Herbig Ae/Be stars as well as young substellar objects. We could potentially measure the ages of deeply embedded protostars. With a large sample of ages, we can probe how the circumstellar environment evolves, from the envelope to the disk to planet formation.

## Acknowledgments

We acknowledge the support of the NASA Infrared Telescope Facility, which is operated by the University of Hawaii under contract 80HQTR24DA010 with the National Aeronautics and Space Administration. We are grateful for the professional assistance from Dave Griep, Brian Cabreira, Greg Engh, Tony Matulonis, and Bernie Walp. This research has made use of the SIMBAD database, operated at CDS, Strasbourg, France, and NASA's Astrophysics Data System. This publication makes use of data products from the Two Micron All Sky Survey, which is a joint project of the University of Massachusetts and the Infrared Processing and Analysis Center/California Institute of Technology, funded by the National Aeronautics and Space Administration and the National Science Foundation. This research has made use of NASA's Astrophysics Data System, operated by the Smithsonian Astrophysical Observatory under NASA Cooperative Agreement 80NSSC21M0056. This publication makes use of data products from the Wide-field Infrared Survey Explorer, which is a joint project of the University of California, Los Angeles, and the Jet Propulsion Laboratory/California Institute of Technology, funded by the National Aeronautics and Space Administration.

*Facility:* IRTF.

## ORCID iDs

Michael Connelley https://orcid.org/0000-0002-8293-1428
Christian Flores https://orcid.org/0000-0002-8591-472X
Bo Reipurth https://orcid.org/0000-0001-8174-1932